\documentclass[12pt,%
 aip,jcp,
groupedaddress,
 amsmath,amssymb,
 reprint,%
 longbibliography,
floatfix,
]{revtex4-1}

\usepackage{graphicx}
\usepackage{amssymb}
\usepackage{xcolor}
\usepackage{dcolumn}
\usepackage{bm}
\usepackage{lipsum}

\usepackage[utf8]{inputenc}
\usepackage[T1]{fontenc}
\usepackage{mathptmx}
\usepackage{etoolbox}
\usepackage{tabularx}
\usepackage{float}
\usepackage{caption}
\usepackage{subcaption}
\usepackage[normalem]{ulem}
\renewcommand{\vec}[1]{\textit{\textbf{#1}}}
\newcommand{\q}{\vec{q}}
\newcommand{\x}{\vec{x}}
\newcommand{\p}{\vec{p}}
\renewcommand{\S}{\vec{S}}

\newcommand{\refeq}[1]{Eq.~\eqref{#1}}
\newcommand{\refsec}[1]{Sec.~\ref{#1}}
\newcommand{\reffig}[1]{Fig.~\ref{#1}}
\newcommand{\reftable}[1]{Table~\ref{#1}}
\newcommand{\stateP}{\mathbb{P}}
\newcommand{\stateR}{\mathbb{R}}
\newcolumntype{K}{>{\centering\arraybackslash}p}

\usepackage[version=4]{mhchem}
\newcommand{\red}[1]{{\color{red}{#1}}{}}

\DeclareMathOperator{\Tr}{Tr}
\DeclareMathOperator{\sgn}{sgn}
\usepackage{braket}
\newcommand{\ketbra}[2]{\ket{#1}\!\bra{#2}}
\usepackage{comment}
\usepackage{hyperref}
\usepackage{soul}

\makeatletter
\def\@email#1#2{%
 \endgroup
 \patchcmd{\titleblock@produce}
  {\frontmatter@RRAPformat}
  {\frontmatter@RRAPformat{\produce@RRAP{*#1\href{mailto:#2}{#2}}}\frontmatter@RRAPformat}
  {}{}
}%
\makeatother
\begin{document}


\title 
{Nonadiabatic rare events from transition-path sampling of MASH trajectories}
\author{Danial Ghamari}
\author{Jeremy O. Richardson}
\email{jeremy.richardson@phys.chem.ethz.ch}
\affiliation{Institute of Molecular Physical Science, ETH Zurich, 8093 Zurich, Switzerland}

\date{\today}

\begin{abstract}
Rare nonadiabatic reactions are a key component of many important molecular processes but are challenging to capture with direct dynamical simulations. In this paper, we combine our recently developed mapping approach to surface hopping (MASH) with transition-path sampling to create a framework to efficiently simulate these rare events.
This is possible because MASH trajectories are Markovian, time-reversible and obey Liouville's theorem.
The combined approach generates nonadiabatic reactive pathways without biasing the underlying dynamics. The resulting ensemble allows for a detailed analysis of reaction mechanisms and the unraveling of statistical and dynamical properties, including rate constants.
We apply the method to study a spin--boson model in thermal equilibrium over a wide range of diabatic coupling strengths.
Our results demonstrate how this approach provides a practical and systematic tool for investigating rare nonadiabatic processes, potentially beyond the reach of brute-force simulations.
\end{abstract}

\maketitle

\section{Introduction}

When a molecule or material is excited by the absorption of light, it can induce dynamical processes relevant to physics, chemistry, and biology.\cite{NonadiabaticBook}
Theoretical descriptions of these photoinduced phenomena require the time evolution of coupled electronic and nuclear degrees of freedom (DoF) after an initial nonadiabatic excitation from the ground state to a higher potential energy surface (PES). Even though these dynamics could, in principle, be described with a full quantum-mechanical treatment, it remains practically impossible due to the computational demand that scales exponentially with the dimension of the system's Hilbert space. 
Fortunately, the relatively high mass of the nuclei justifies treating them as classical particles with well-defined positions and momenta, as long as the electronic DoF remain fully quantum mechanical. This idea has motivated the development of a large class of mixed quantum--classical methods,\cite{Stock2005nonadiabatic} of which Tully's fewest-switches surface hopping (FSSH) is one of the most popular and well-known approaches.\cite{Tully1990hopping,Subotnik2016review,Barbatti2014newtonX,Mai2018SHARC}

FSSH introduces a stochastic hopping between states to generalize classical molecular dynamics to nonadiabatic processes. Despite its popularity, however, the method lacks a rigorous derivation and a number of major challenges have been identified.\cite{Granucci2007FSSH} 
For instance, as FSSH trajectories evolve, the electronic DoF and the active state do not necessarily remain consistent.\cite{Fang1999}
Moreover, the trajectories do not satisfy microscopic reversibility,\cite{HammesSchiffer1995FSSH} and their ensemble averages are not guaranteed to relax to the correct Boltzmann equilibrium.\cite{Schmidt2008equilibrium}
As such, to utilize FSSH for the study of nonadiabatic reactions and rates, one is typically forced to rely on long-time simulations of population transfer 
and cannot benefit directly from more efficient rare-event theories.\cite{PetersBook} 
While some photoinduced reactions occur on a femtosecond timescale, others may take nanoseconds or longer due to energetic or entropic barriers on the excited PES\@. 
There is thus a significant challenge due to the separation of timescales between the fast intrinsic dynamics of the molecular vibrations and the slow reaction process.
Modern attempts to tackle this problem have been based around constructing cheap machine-learned PESs to make the direct simulation of the rare events feasible.\cite{Westermayr2020perspective,Li2022ML}
As useful as these approaches are, they do not really get to the heart of the problem, which is that the majority of the computer time is spent simulating uninteresting dynamics, while the important reactive transitions occur only rarely.

Such challenges are not particular to the simulation of nonadiabatic dynamics, and
in the field of
classical molecular dynamics (on a single Born--Oppenheimer PES), 
many enhanced-sampling methods have been developed to address the same difficulty posed by the separation of timescales.\cite{PetersBook} 
In particular, transition-path sampling (TPS), developed by Dellago et al.,\cite{Dellago1998,Dellago1998_2} is a method that captures rare events 
without needing to parameterize a reaction coordinate or even a set of good collective variables (CVs). 
The goal in TPS is to utilize Monte Carlo sampling to focus the computational power solely on generating reactive trajectories, and avoiding wasting time sampling paths that do not contribute to the reaction under study. In more than 20 years since the development of TPS and its variant, forward-flux sampling,\cite{Allen2005,Allen2006} these approaches have demonstrated great success in simulating rare events.\cite{Juraszek2006,Escobedo2009,Sosso2016,Hussain2020,Bolhuis2021,Royall2024} Even though most of these applications have been for systems evolving (classically) on a single PES, attempts were also made to integrate path-sampling schemes with FSSH to study nonadiabatic reactions. A particular complication is the lack of microscopic reversibility along an FSSH trajectory, which is a key component of TPS\@. Sherman and Corcelli \cite{Sherman2016} utilized a reweighting scheme, originally proposed by Hammes-Schiffer and Tully, \cite{HammesSchiffer1995FSSH} 
to circumvent this difficulty. However, this importance-sampling approach 
can significantly reduce the sampling efficiency, whenever the approximate hopping probabilities used to generate the trial backwards trajectory deviate strongly from the ``correct'' FSSH probabilities. 
Additionally, Reiner et al.\cite{Reiner2023} developed a method based on combining FSSH with forward-flux sampling. In contrast to TPS, this approach does not impose any requirement on the microscopic reversibility of dynamics or knowledge of the equilibrium probability distribution in the reactant. However, the efficiency of forward-flux sampling is highly sensitive to the choice of the reaction coordinate, and the location of interfaces in that space.\cite{Escobedo2009} Nevertheless, the application of this method to two model systems, one with an avoided crossing and another with a conical intersection, obtained good overall agreement with brute-force FSSH calculations, while increasing the efficiency of simulation to varying degrees.\cite{Reiner2023}

Recently, our group has developed an alternative quantum--classical method for simulating nonadiabatic dynamics called the mapping approach to surface hopping (MASH).\cite{MASH,MASHreview} It was designed to combine the practicality offered by FSSH with the rigor of linearized semiclassical mapping approaches.\cite{Meyer1979nonadiabatic,Stock1997mapping,Sun1998mapping,Wang1999mapping,Miller2016Faraday,spinmap,multispin}
Like FSSH, MASH is computationally efficient, and can be applied in conjunction with on-the-fly ab initio electronic-structure theory to simulate photochemistry.\cite{Mannouch2024MASH,cyclobutanone}
However, unlike FSSH, MASH's equations of motion are deterministic and time-reversible.\cite{MASHEOM}
Reference~\onlinecite{MASHrates} showed that this makes it possible to employ the Bennett--Chandler reactive-flux approach\cite{MolSim,ChandlerGreen} within the MASH formalism to efficiently calculate the rate constant of nonadiabatic reactions, as long as the reaction coordinate and the transition region are sufficiently easy to identify.
In addition to offering this significant computational advantage, the MASH predictions were more reliable than those of FSSH due to the fact that it guarantees consistency between the most populated electronic state and the active surface.  In particular, MASH was able to correctly capture the known behavior of Marcus theory, where FSSH predicts an incorrect scaling unless complicated decoherence corrections are applied.\cite{Landry2011hopping,Landry2012hopping}

In general, however, it is not always so easy to identify a good reaction coordinate in order to apply the Bennett--Chandler approach. 
For this reason, it is worth investigating the possibility of using more powerful enhanced-sampling methods such as TPS\@.
In this work, we thus propose a nonadiabatic framework called MASH-TPS, which samples an ensemble of reactive MASH trajectories and allows for an efficient calculation of the nonadiabatic reaction rate.
In addition, this ensemble can be used to interpret the mechanism and identify the most important collective variables in a reaction. 
After introducing the formalism behind both the MASH and TPS approaches, we apply the combined method to study the nonadiabatic reaction rate in a spin--boson model over a range of coupling strengths from the adiabatic to the nonadiabatic limit.

\section{Enhanced sampling of nonadiabatic dynamics}
In the following, we first provide a brief description of how MASH can be used to simulate the nonadiabatic dynamics of a two-level quantum system coupled to multiple classical degrees of freedom.  In particular, we point out differences between MASH and the commonly-used FSSH formalism. 
Then, we introduce the general framework of TPS and describe how it can be extended to obtain an ensemble of reactive MASH trajectories. 
Finally, we explain how to calculate the nonadiabatic rate constant from 
the transition-path ensemble.


\subsection{Nonadiabatic dynamics with MASH}\label{SEC:MASH}
In mixed quantum--classical dynamics, the coordinates and momenta of $f$ nuclear degrees of freedom are represented as classical variables, $\q =\{q_1,\dots,q_{f}\}$ and $\p=\{p_1,\dots,p_f\}$. 
The electronic degrees of freedom are treated quantum mechanically in a basis of adiabatic states, $\ket{\Phi_\pm(\q)}$, which diagonalize the potential operator (also known as the electronic Hamiltonian), $\hat{V}(\bm{q})=\bar{V}(\q)\hat{I} + V_z(\q)\hat{\sigma}_z(\q)$, where $\hat{\sigma}_z(\q)=\ketbra{\Phi_+}{\Phi_+} - \ketbra{\Phi_-}{\Phi_-}$ is the Pauli matrix at the nuclear configuration $\q$ and $\hat{I}$ is the identity operator.
Here, we have assumed that the dynamics of interest can be described using a truncated electronic Hilbert space of two adiabatic states.

In this representation,
the general {nonadiabatic} Hamiltonian that couples the nuclear and electronic DoF is written as 
\begin{equation}
    \hat{H}(\q,\p) = \sum^{f}_{j=1}\frac{p_j^2}{2m_j} + \hat{V}(\q) ,
\end{equation}
where $m_j$ is the mass associated with the $j$-th degree of freedom.
The nuclei follow classical equations of motion,
\begin{align}
    \dot{q}_j &= p_j / m_j, & \dot{p}_j &= F_j,
\end{align}
where the definition of the force components, $F_j$, will be discussed later.

Following ideas of the spin-mapping approach,\cite{spinmap}
MASH employs an isomorphism between a two-level system and a spin-$\frac{1}{2}$ particle, which maps the instantaneous electronic wavefunction $|\Psi\rangle = c_+|\Phi_+(\q)\rangle + c_-|\Phi_-(\q)\rangle$ onto a (normalized) spin vector $\S=(S_x,S_y,S_z)$ on a Bloch sphere. The relations between the coefficients $c_\pm$ of the wavefunction and the spin variables are
\begin{subequations} 
\begin{align}
    S_x &= 2\,\text{Re}\,[c_+c_-^*],\\
    S_y &= 2\,\text{Im}\,[c_+c_-^*],\\
    S_z &= |c_+|^2-|c_-|^2 .
\end{align}
\end{subequations}
In this form, the Schrödinger equation governing the electronic DoF as the system moves along the nuclear trajectory is 
\begin{subequations} \label{EQ:MASH_spin_eof}
\begin{align}
    \dot{S}_x &= \sum_j\frac{2d_j(\q)p_j}{m_j} S_z - \frac{2}{\hbar}V_z(\q)S_y,\\
    \dot{S}_y &= \frac{2}{\hbar}V_z(\q)S_x,\\
    \dot{S}_z &= -\sum_j\frac{2d_j(\q)p_j}{m_j}S_x.
\end{align}
\end{subequations}
Here, $d_{j}(\bm{q}) = \langle\Phi_+(\q)|\frac{\partial}{\partial q_j}\Phi_-(\q)\rangle$ is the $j$-th element of the nonadiabatic coupling vector, which accounts for the change of basis along the trajectory.\cite{NonadiabaticBook,MASHreview} 

MASH is inspired by the FSSH approach,\cite{Tully1990hopping} in which the nuclei move according to the force of the active adiabatic state.
However, whereas FSSH stochastically hops between active states with probabilities determined by the electronic coefficients,
MASH hops deterministically according to the value of $\sgn(S_z)$, where $\text{sgn}(S_z) = -1,0,1$ for $S_z<0$, $S_z=0$, and $S_z>0$.\cite{MASH}
The nuclei thus move according to the force of the upper state when the spin vector is in the northern hemisphere and hop down to the lower state when the spin vector crosses the equator of the Bloch sphere.
In this way, MASH avoids the inconsistency problem of FSSH, in which the active state may differ from the most populated electronic state leading to significant errors.\cite{Fang1999,Granucci2007FSSH,Landry2011hopping,MASHrates}
The force in MASH is thus defined as
\begin{equation}\label{EQ:mash_force}
    F_j = -\frac{\partial\bar{V}}{\partial q_j} - \frac{\partial V_{z}}{\partial q_j} \sgn(S_z)+ 4V_z(\q)\,d_j(\q)\,S_x\,\delta(S_z).
\end{equation}
The final term applies an impulse when a hop from one surface to another occurs and rescales the momenta (along the direction of the nonadiabatic coupling vector) to ensure conservation of the energy function
\begin{equation}
    E(\q,\p,\S) = \sum^{f}_{j=1}\frac{p_j^2}{2m_j} + \bar{V}(\q) + V_z(\q)\text{sgn}(S_z) .
\end{equation}
In cases when a trajectory has insufficient energy to hop up (called a frustrated hop), the momentum is reflected and the spin vector remains in the lower hemisphere.\cite{MASH}
Importantly,
unlike FSSH,
whose active state depends on the history of the trajectory, 
\cite{HammesSchiffer1995FSSH}\footnote{Hammes-Schiffer and Tully used a phase space consisting of $\bm{q}$, $\bm{p}$ and the active state label.  If instead one additionally includes $c_\pm$, the FSSH trajectories can be thought of as Markovian.  However, unlike MASH, where trajectories are initialized in a hemisphere, FSSH trajectories are defined to start at a pole of the Bloch sphere, which is of measure 0 such that (almost) all trajectories generated by TPS will be rejected.}
the MASH equations of motion are not only time-translationally invariant but also time-reversible.\cite{MASHEOM} 
This property allows MASH to be naturally incorporated into the TPS formalism, as we shall explore in this work.




We must also specify the procedure to calculate observables from ensembles of MASH trajectories. Here, we only focus on observables defined in terms of the adiabatic populations and nuclear configurations, although we note that MASH additionally provides rigorous prescriptions for cases involving electronic coherences.\cite{Mannouch2024coherence,MASHcoh} 
In particular, the population projection operators $\ketbra{\Phi_\pm}{\Phi_\pm}$ are represented by the Heaviside step functions $h(\pm S_z)$ while nuclear operators are represented by their classical counterparts.

In the MASH formalism, correlation functions take the form of an ensemble average
\begin{equation}
    C_{AB}(t) = \big\langle A(\q,\p,\S)\,B(\q(t),\p(t),\S(t))  \big\rangle ,
\end{equation}
where $B(\q,\p,\S)$ is the MASH representation of a measurement operator.\cite{MASHreview}
Here, $\langle \cdots\rangle$ is an integral over the phase space variables
\begin{align}\label{EQ:mash_correlation}
    \langle \cdots\rangle = \iiint \mathrm{d}\q\,\mathrm{d}\p\,\mathrm{d}\S \;\rho_0(\q,\p,\bm{S}) \, W(\bm{S})\, \cdots ,
\end{align}
where $\rho_0(\q,\p,\bm{S})$ and $A(\q,\p,\S)$ together define the initial distribution but are defined as two terms for notational convenience.\footnote{Classical distributions are used throughout this work rather than Wigner functions so as to avoid zero-point energy leakage and take advantage of the statistical properties of classical equilibrium distributions combined with classical dynamics.}
The term $W(\bm{S})$ is a positive weighting factor that was constructed to ensure that the MASH correlation function matches the quantum-mechanical electronic evolution along a pre-specified nuclear path.\cite{MASH,MASHreview} 
For all the cases relevant to this work,
the weighting factor is $W(\S) = 2|S_z|$.
Note that all the ensemble averages below follow the formulation specified in \refeq{EQ:mash_correlation}, i.e., including $\rho_0$ and $W(\S)$. The integration over the $\S$ variable is defined by
\begin{equation}
    \int \mathrm{d}\S\cdots = \frac{1}{2\pi}\int_0^\pi \sin\theta\,\mathrm{d}\theta\int_0^{2\pi} \mathrm{d}\varphi\cdots ,
\end{equation}
where $S_x = \sin\theta\cos\varphi$, $S_y = \sin\theta\sin\varphi$, and $S_z = \cos\theta$.
In the standard MASH approach, this integral is evaluated by Monte Carlo sampling of $\q$, $\p$, $\cos\theta$ and $\varphi$, typically relying on uniform sampling of the Bloch sphere followed by reweighting. 
MASH trajectories conserve energy, and thus correctly preserve an equilibrium ensemble such as $\rho_0\propto\exp[-\beta E(\q,\p,\S)]$.
The dynamics also preserve phase-space volume according to Liouville's theorem, $\mathrm{d}\q\,\mathrm{d}\p\,\mathrm{d}\S=\mathrm{d}\q(t)\,\mathrm{d}\p(t)\,\mathrm{d}\S(t)$.\cite{MASH}
Although MASH does not rigorously obey $C_{AB}(t)=C_{BA}(-t)$ [because of the existence of the weighting function $W(\S)$], the trajectories themselves are time-reversible.  
This is sufficient to enable the direct use of the TPS technique.

\subsection{Transition-path sampling}\label{SEC:TPS}
In MASH, a microstate is completely defined by a point in the extended phase space $\x=\{\q,\p,\S\}$ and 
a trajectory can be denoted as an ordered sequence of states visited along the path\cite{Dellago2006TPS,TuckermanBook}
\begin{equation}\label{EQ:traj}
    X(T) = \left\{\x_0,\x_{1},\x_{2},\dots,\x_{N}\right\}.
\end{equation}
Here, the trajectory has been discretized using timesteps of $\delta t$, such that the total time is $T = N\delta t$, and $\x_i$ denotes a point in the phase space occupied at the \textit{i}-th step. 
Assuming Markovianity,
the probability of observing this path is given by
\begin{equation}\label{EQ:general_path_probability}
    \mathcal{P}[X(T)] = \rho_0(\x_0)\prod^{N-1}_{i=0} P(\x_i\to \x_{i+1}) ,
\end{equation}
where $\rho_0(\x_0)$ is the unconstrained probability distribution of the starting point, 
and $P(\x_i\to \x_{i+1})$ is the transition probability along the trajectory. In the deterministic dynamics of MASH, the latter takes the form of Dirac's delta function $\delta(\x_{i+1}-\phi_{\delta t}(\x_i))$, where $\phi_{\delta t}(\x)$ is the flow map\cite{LeimkuhlerBook} determined by the finite-time integrator of MASH.\cite{MASHEOM} Bear in mind that this form is only applicable in the absence of an external source of stochastic noise, such as a heat bath. In this case, the MASH equations include a Langevin friction with random forces, which we will discuss in 
\refsec{SEC:results}.  However, as long as the friction is Markovian, one can still express the probability of the path as in \refeq{EQ:general_path_probability}, albeit with Gaussian transition probabilities, which cancel out in the final expressions.

The probability distribution that characterizes the ensemble of reactive paths which connect the reactant $\stateR$ and product $\stateP$ is\cite{TuckermanBook,Dellago1998}
\begin{equation}
    \mathcal{P}_{\text{react.}}[X] = \frac{1}{\mathcal{Z}_{\text{react.}}}h_\stateR(\x_0)h_\stateP(\x_N)\mathcal{P}[X]\label{EQ:reactive_traj_probability} ,
\end{equation}
where the functions $h_\stateR(\x)$ and $h_\stateP(\x)$ return unity if $\x$ is located in the reactant/product basin and $0$ otherwise, and the partition function $\mathcal{Z}_{\text{react.}}$ ensures the probability is normalized in the subspace of reactive paths, $\int \mathcal{D}[X] \mathcal{P}_{\text{react.}}[X] = 1$. 
We note that the definition of the functions $\stateR$ and $\stateP$ typically relies on a CV (or a collection of CVs). However, this is the only part of algorithm that depends on user-selected CVs. 
Moreover, it is often much easier to identify an adequate CV for the definition of the two states, than parameterizing a good reaction coordinate for enhancing the sampling. This is the main advantage of TPS over alternative rare-event methods.

In order to initilize the Monte Carlo (MC) process for TPS, we require a single arbitrary reactive path.
Here, one often relies on direct simulations to capture a rare transition between the two metastable states; however, especially in high-dimensional problems where entropic effects make transitions very unlikely, this arduous task could be achieved with more efficient approaches, such as metadynamics.\cite{Laio2002} In the present study, generating this initial path was not problematic as the barriers were low enough
that direct simulations from the reactant were sufficient to generate a few reactive trajectories.

Once the (old) path $X^\text{old}$ is obtained, 
various techniques 
exist to generate a new trial pathway $X^\text{new}$. Among them, the original suggestions of \textit{shooting} and \textit{shifting} moves are simple and efficient.\cite{TPSreview}  We have thus adopted them in this paper and will discuss their generalization to MASH dynamics in Secs.~\ref{SEC:shooting} and \ref{SEC:shifting}.
After generating this trial pathway, we decide whether to accept or reject it by invoking the Metropolis criterion with the acceptance probability:
\begin{align}\label{EQ:acceptance_prob}
    \nonumber &\mathcal{P}_{\text{acc.}}[X^{\text{old}}\to X^{\text{new}}] = \\
    &\min\left\{1,\,\frac{\mathcal{P}_{\text{react.}}[X^{\text{new}}]\,\mathcal{P}_{\text{gen.}}[X^{\text{new}}\to X^{\text{old}}]}{\mathcal{P}_{\text{react.}}[X^{\text{old}}]\,\mathcal{P}_{\text{gen.}}[X^{\text{old}}\to X^{\text{new}}]}\right\} .
 \end{align}
The form of the generation probability $\mathcal{P}_{\text{gen.}}[X^{\text{old}}\to X^{\text{new}}]$ depends on the trial move utilized, as described in Secs.~\ref{SEC:shooting} and \ref{SEC:shifting}. 
This criterion in \refeq{EQ:acceptance_prob} ensures detailed balance in the path space, $\mathcal{P}_\mathrm{react.}[X^\mathrm{old}]{\mathcal{P}_{gen.}[X^\mathrm{old}\rightarrow X^\mathrm{new}]}\mathcal{P}_\mathrm{acc}[X^\mathrm{old}\rightarrow X^\mathrm{new}] = \mathcal{P}_\mathrm{react.}[X^\mathrm{new}]\mathcal{P}_\mathrm{gen.}[X^\mathrm{new}\rightarrow X^\mathrm{old}]\mathcal{P}_\mathrm{acc.}[X^\mathrm{new}\rightarrow X^\mathrm{old}]$. 
Finally, by iteratively generating new paths using the trial moves, 
we obtain an ensemble of reactive pathways between the two states with the probability \refeq{EQ:reactive_traj_probability}.

\subsubsection{Shooting moves}\label{SEC:shooting}
In a shooting move, we start by randomly choosing (with a discrete uniform distribution) a state $\x^\text{old}_n$ along the old path from the transition region (between the reactant and product states). Typically, in applications of TPS to Newtonian dynamics, a new state is generated by adding a perturbation in the form of 
random Gaussian noise $\delta\p$ to the momentum of the old state $\p_n^\text{new}=\p_n^\text{old}+\delta\p$. 
The noise is taken to have zero mean and a variance empirically adjusted to optimize the sampling efficiency. 

In MASH dynamics, however, the momentum perturbation alone may not be sufficient. 
Therefore, we also introduce shooting moves for the spin vector $\S$ using the following procedure. We start by finding two orthogonal vectors to $\S$, e.g., $\S' = \alpha(1/S_x,-1/S_y,0)$ with $\alpha=|S_xS_y|/\sqrt{(S^2_x+S_y^2)}$ and $\S''=\S\times\S'$. Next, we generate a random vector, 
$\bm{\Gamma} = \bm{S}' \cos\eta + \bm{S}'' \sin\eta$, where $\eta\in[0,2\pi)$ is sampled uniformly.
Next, we rotate $\S$ around $\bm{\Gamma}$ with the random angle $\chi$ (chosen from a Gaussian distribution centred at 0) 
\begin{equation}
    \S^\text{new}_n = \exp\left[\chi\textbf{W}(\bm{\Gamma})\right]\cdot\S_n^\text{old}, 
\end{equation}
where
\begin{align}
   \textbf{W}(\bm{\Gamma}) &= \left(\begin{array}{ccc}
        0 &  -\Gamma_z&  \Gamma_y\\
        \Gamma_z & 0 &  -\Gamma_x\\
        -\Gamma_y &\Gamma_x &  0
    \end{array}\right).
\end{align}
Like the momentum shooting, this transformation ensures that the perturbation is symmetric, i.e., the probability to go from $\bm{S}^\text{old}_n$ to $\bm{S}^\text{new}_n$ is the same as the probability to go from $\bm{S}^\text{new}_n$ to $\bm{S}^\text{old}_n$.

In the next step of shooting, we initialize two MASH simulations from the new state, one forwards and one backwards in time. The procedure for the former is a straightforward integration of MASH equations for the interval $[n\delta t, T]$. For the latter simulation however, according to equations of MASH, we first have to reverse $\p_n^{\text{new}}\to-\p_n^{\text{new}}$, and $(S_y)_n^\text{new}\to(-S_y)_n^\text{new}$, and then integrate until $t=0$ is reached.\footnote{Alternatively, one can simply use the same integrator with a negative value of $\delta t$, taking advantage of our recently-derived reversible integrators.\cite{MASHEOM}}
Finally, by applying the same reversal procedure on the backward trajectory and appending it to the beginning of the forward segment, we obtain a new trial trajectory $X^{\text{new}}(T)$. The generation probability of this path is given by
\begin{align}
   \nonumber \mathcal{P}_{\text{gen.}}[X^{\text{old}}\to X^{\text{new}}] &= P_{\text{shoot}}(\x_n^\text{old}\to \x_n^\text{new}) 
   \\
   &\quad\times\prod_{i=0}^{N-1}
   P(\x_{i}^\text{new}\rightarrow\x_{i+1}^\text{new}),
\end{align}
where $P_\text{shoot}$ 
is the probability of singling out the state in the transition region of the path and generating the new state (including the choice of the shooting point $n$ and is thus inversely proportional to the number of points in the transition region, $N_\text{shoot}$).
Inserting this expression in \refeq{EQ:acceptance_prob} and bearing in mind that every old path must be reactive, the Metropolis criterion can be simplified to
\begin{equation}\label{EQ:acceptance_prob_shooting}
    \mathcal{P}_{\text{acc.}}[X^{\text{old}}\to X^{\text{new}}] = h_\stateR(\x^{\text{new}}_0)h_\stateP(\x^{\text{new}}_N)\min\left\{1,\frac{N_\text{shoot}^\text{old}\rho_0(\x^{\text{new}}_0)}{N_\text{shoot}^\text{new}\rho_0(\x^{\text{old}}_0)}\right\}.
\end{equation}

\subsubsection{Shifting moves}\label{SEC:shifting}
A particularly computationally efficient method to generate new pathways is by time-translation of the initial conditions of previously generated paths. 
For instance, the new trial path could be selected as $X^\text{new}(T) = \{\x^\text{old}_{L},\dots,\x^\text{old}_{N+L}\}$ or $X^\text{new}(T) = \{\x^\text{old}_{-L},\dots,\x^\text{old}_{N-L}\}$, where the old path has been extended by $L$ steps in either the forward or backward direction. Provided that 
the direction is chosen at random, the 
generation probabilities are symmetric and the acceptance probability of such a move is
also given by \refeq{EQ:acceptance_prob_shooting}.

It is generally suggested to attempt shifting moves much more frequently than the computationally more expensive shooting ones. Even though shifting does not result in significantly different new paths, it provides statistical refinement for variables and correlation functions that one obtains from analyzing the TPS ensemble. 

\subsection{Rate constant}\label{SEC:rate_const}
In a reaction process, the reactive flux $k(t)$ is one of the most important quantities to determine. For a dynamical system with two metastable states, this function is related to the time correlation function $C(t)$ through\cite{ChandlerGreen}
\begin{align}\label{EQ:kinetic_rate_1}
    k(t) &= \frac{\mathrm{d}}{\mathrm{d}t}C(t), & C(t) &= \frac{\langle h_\stateR(0) \, {h}_\stateP (t) \rangle}{\langle h_\stateR\rangle},
\end{align}
where we use a simplified notation $h_\stateP(t) \equiv h_\stateP(\x_i)$ for $t=i\delta t$.
Here, the ensemble average is defined in \refeq{EQ:mash_correlation}. 
Therefore, $k(t)$ signifies the rate at which the fraction of reactive trajectories (which start in $\stateR$ and end in $\stateP$) builds up in time.

It is expected that $C(t)$ approaches its asymptotic value exponentially\cite{ChandlerGreen}
\begin{equation}
    C(t)\sim \braket{h_\stateP}_\infty (1-\mathrm{e}^{-t/\tau_\text{rxn}}),
\end{equation}
where $\braket{h_\stateP}_\infty$ is the product population in the long-time limit and $\tau_\text{rxn} = [k_{\stateR\stateP}+k_{\stateP\stateR}]^{-1}$ is the reaction time with $k_{\stateR\stateP}$ ($k_{\stateP\stateR}$) being the forward (backward) rate constant that we wish to compute. By definition, in rare reactions, there exists a gap between the slow and fast modes of the dynamics. Consequently, there exists a regime 
longer than the characteristic molecular timescale, $\tau_\text{mol}$, but shorter than the overall reaction time
where 
\begin{equation}
    k(t)\approx k_{\stateR\stateP} \quad\text{for}\quad \tau_{\text{mol}}\ll t \ll\tau_{\text{rxn}}.
\end{equation}
Therefore, in this regime, the reactive flux, $k(t)$, displays a plateau whose value is the forward rate constant. Hereafter, wherever the rate constant is mentioned, we refer to this quantity.

Of course, calculating the rate constant as the direct time derivative of $C(t)$ is computationally very difficult, due to the rarity of the reaction event. 
Fortunately, one can circumvent this difficulty through utilizing the TPS framework.\cite{Dellago1999} 
%
We start by factorizing $C(t)$ from \refeq{EQ:kinetic_rate_1}: 
\begin{align}\label{EQ:Ct_MASH_factor1}
   \nonumber  C(t) =& \frac{\langle  h_\stateR (0)\, h_\stateP (t) \rangle}{\langle h_\stateR (0)\, h_\stateP ({T'})\rangle} \cdot \frac{\langle h_\stateR (0) \, h_\stateP ({T'})\rangle}{\langle h_\stateR\rangle}\\ =& \frac{\langle  h_\stateR(0) \,h_\stateP (t) \rangle}{\langle h_\stateR(0)\, h_\stateP ({T'})\rangle} \, C(T').
\end{align}
We then define the indicator function
\begin{equation}
    H_\mathbb{P}(X) = \max_{t\leq T}\,h_\stateP(t) ,
\end{equation}
which equals 1 if at any time the trajectory $X(T)$ visits the product state (regardless of where in the phase space it ends), and 0 otherwise. For $t\in[0,T]$, the equality $h_\stateP(t)H_\mathbb{P}(X) = h_\stateP(t)$ allows us to rewrite \refeq{EQ:Ct_MASH_factor1} as:
\begin{align}
   C(t) = &\frac{\langle  h_\stateR(0) h_\stateP (t)H_\stateP (X) \rangle}{\langle  h_\stateR(0) H_\stateP (X) \rangle}
   \cdot
   \frac{\langle  h_\stateR(0) H_\stateP (X) \rangle}{\langle h_\stateR(0)  h_\stateP ({T'})H_\stateP (X)\rangle} \, C(T').
\end{align}
This expression can be written concisely as
\begin{align}\label{EQ:final_Ct}
    C(t)= \frac{\langle h_\stateP(t)\rangle_{\text{TPS}}}{\langle h_\stateP({T'})\rangle_{\text{TPS}}} \, C(T'),
\end{align}
where $\langle\dots\rangle_\text{TPS}$ indicates averaging over the TPS ensemble [weighted by $\rho_0$ and $W(\S)$] of all trajectories of length $T$ that start in the reactant and by the time $T$, have passed through the product's boundary at least once. Using this factorization, the ardeous task of evaluating $C(t)$ for a range of times has been replaced by one TPS ensemble average scaled by the value of $C(T')$ evaluated only for a single reference time, $T'$, whose value can be taken much smaller than $T$. 
The value of $C(T')$
can be evaluated using 
a similar TPS algorithm\cite{Dellago1999} and the Weighted Histogram Analysis Method (WHAM). \cite{WHAM}

In particular, $C(T')$ can be calculated using a simple one-dimensional reaction coordinate similar to a standard free-energy calculation. The space spanned by the reaction coordinate is divided into $M$ overlapping bins. Then, for each bin, we initiate a TPS simulation where trajectories of length $T'$ starting from the reactant are accepted only if their endpoints fall in the respective bin. Using the collected endpoints, we construct a probability distribution in the reaction-coordinate space. Next, WHAM is utilized to connect the overlapping histograms into a complete, normalized conditional probability distribution. $C(T')$ is given by the integral over this probability distribution in the product region.
Finally, the correlation function is reconstructed from \refeq{EQ:final_Ct} for $t\in[0,T]$ and the rate constant can be extracted from its numerical time-derivative in the plateau region.
This TPS procedure can be far more computationally efficient than evaluating $C(t)$ directly using \refeq{EQ:kinetic_rate_1}, especially for slow reactions.\cite{Dellago1999}
Although in principle the TPS rate calculation should recover exactly the same result as a converged brute-force simulation, in practice, the calculation of $C(T')$ is hampered by a compromise between accuracy and efficiency in selecting the bin size and by the reliability of the WHAM procedure.
Future work will explore whether alternative approaches such as MBAR (multistate Bennett acceptance ratio)\cite{Shirts2008} can be used to achieve more accurate results.

\section{Results} \label{SEC:results}
For the application of MASH-TPS, we study the symmetric spin--boson model, which is often taken as a prototype for a wide variety of chemical and physical processes, e.g., electron transfer and polaron dynamics.\cite{Leggett1987spinboson,Weiss,Marcus1985review} In the Brownian-oscillator formulation, the diabatic potential acting on the mass-weighted solvent polarization coordinate, $R$ (which here plays the role of $\sqrt{m_j}q_j$) is given by 
\begin{align}
   \hat{V}(R) &= \left(\begin{array}{cc}
        V_a(R) &  \Delta\\
        \Delta & V_b(R)
    \end{array}\right),
\end{align}
with $V_{a/b}(R)$ being two harmonic surfaces defined as:
\begin{equation}\label{EQ:Spin_boson_diabatic_eq}
    V_{a/b}(R) = \frac{1}{2} \Omega_0^2\left(R\pm \sqrt{\frac{\Lambda}{2\Omega_0^2}}\right)^2 .
\end{equation}
Here, $\Lambda$ is the Marcus reorganization energy, $\Omega_0$ denotes the characteristic frequency of $R$, and the diabatic coupling $\Delta$ is constant.
In order to employ MASH, we must first transform to an adiabatic basis using $\bar{V}(R)=\frac{1}{2}(V_a+V_b)$, $V_z(R)=\frac{1}{2}\sqrt{(V_a-V_b)^2+4\Delta^2}$ and where $\ket{\Phi_\pm}$ are defined as the eigenvectors of $\hat{V}(R)$.
The gradients can be obtained from the Hellmann--Feynman theorem as
$\frac{\partial \bar{V}}{\partial R} \pm \frac{\partial V_z}{\partial R} = \braket{\Phi_\pm|\frac{\partial \hat{V}}{\partial R}|\Phi_\pm}$ 
and the nonadiabatic coupling vector as
$d_j = - \braket{\Phi_+|\frac{\partial \hat{V}}{\partial R}|\Phi_-} / 2V_z$.\cite{Cederbaum2004CI} 
In order to describe thermally activated reactions,
the solvent coordinate is further coupled to an Ohmic bath with spectral density $J(\omega) = \gamma\omega$. 

\begin{figure}[!t]
    \centering
    \includegraphics[width=1\linewidth]{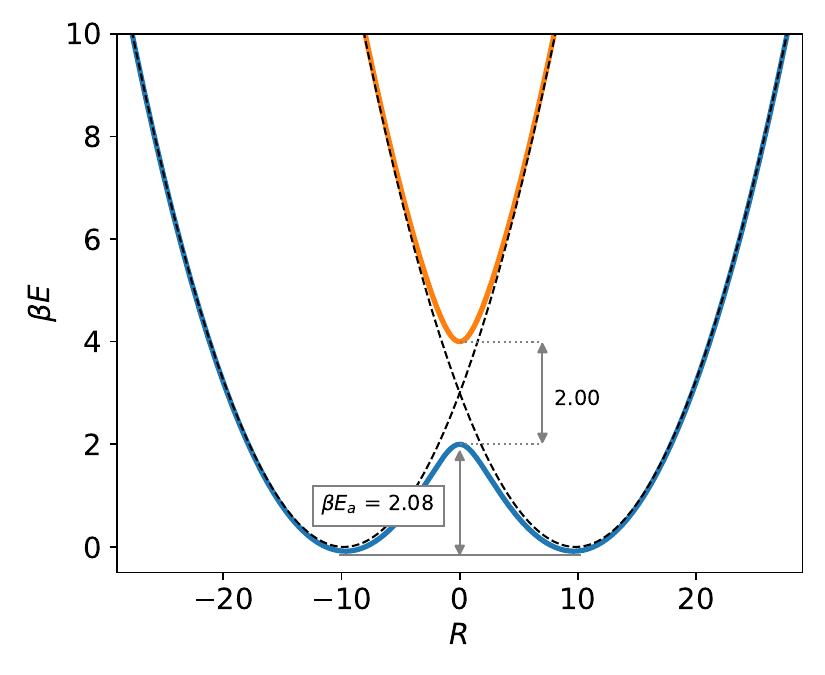}
    \caption{Diabatic (dashed lines) and adiabatic (solid colored lines) potentials of the spin--boson model 
    with $\beta\Delta=1$. The adiabatic energy gap (avoided crossing) at the transition state $R=0$ is $2\Delta$.}
    \label{FIG:SB_PES}
\end{figure}

Recently, Ref.~\onlinecite{MASHrates} compared the numerical calculations of the rate constant for this model obtained via both MASH and FSSH methods and across various diabatic coupling strengths, with an exact quantum description via the hierarchical equations of motion (HEOM).\cite{Tanimura2020HEOM} This study demonstrated that 
MASH captures the rate well across the whole parameter regime (except at low temperatures where quantum nuclear effects such as tunneling would be required),
whereas FSSH fails for weak couplings, as already observed by Landry and Subotnik.\cite{Landry2011hopping}
Consistent with this previous study,
we have set the parameter values: $\beta\Lambda=12$, $\beta\hbar\Omega_0=1/4$, and $\gamma=\Omega_0$, whereas $\beta\Delta$ is varied between the adiabatic and nonadiabatic limiting behaviors. 
The system is illustrated 
in \reffig{FIG:SB_PES}.

Integrating out the bath modes results in an Langevin equation of motion for the solvent polarization coordinate:\cite{Zwanzig,Weiss}
\begin{equation}\label{EQ:spinboson_EOM1}
    \ddot{R} = F(R) - \gamma \dot{R} + \eta(t),
\end{equation}
where $F(R)$ is the standard MASH force [\refeq{EQ:mash_force}] and $\eta(t)$ is a zero-mean stochastic force, defined by $\langle\eta(t)\eta(t')\rangle = 2\gamma\beta^{-1}\delta(t-t')$. 
For the numerical integration of \refeq{EQ:spinboson_EOM1}, we utilize the scheme\cite{Dellago1998_2}
\begin{subequations}\label{EQ:spinboson_EOM2}
    \begin{align}
        R_{i+1} &= R_i +a_1 \dot{R}_i\delta t+ a_2\ddot{R}_i\delta t^2 +\delta R,\\
        \dot{R}_{i+1} &= a_0\dot{R}_i +\left[(a_1-a_2) \ddot{R}_i +a_2\ddot{R} _{i+1}\right] \delta t +\delta \dot{R},
    \end{align}
\end{subequations}
where 
\begin{equation}
    a_0 = \mathrm{e}^{-\gamma \delta t},\quad a_1=\frac{1-a_0}{\gamma \delta t},\quad a_2=\frac{1-a_1}{\gamma \delta t}
.\end{equation}
The random displacements $\delta R$ and $\delta\dot{R}$ at each time step $\delta t$ are taken from a multivariate Gaussian distribution with zero average and a covariance matrix given by
\begin{align}
   &\left(\begin{array}{cc}
        \langle\delta R^2\rangle &  \langle\delta R\,\delta \dot{R}\rangle\\
        \langle\delta \dot{R}\,\delta R\rangle & \langle\delta \dot{R}^2\rangle
    \end{array}\right)\nonumber =
    \\& 
    \frac{\delta t}{\gamma\beta}\left(\begin{array}{cc}
          2-3a_1-a_0a_1& \delta t(\gamma\, a_1)^2\\
         \delta t(\gamma\, a_1)^2& \gamma(1-a_0^2)/\delta t
    \end{array}\right).
\end{align}
It is worth noting that by decreasing the friction $\gamma\to0$, \refeq{EQ:spinboson_EOM2} takes the usual velocity-Verlet form.
The electronic DoF are propagated exactly according to \refeq{EQ:MASH_spin_eof}, 
in steps that alternate with the nuclear propagation.\cite{MASHEOM,Runeson2023MASH,cyclobutanone} This scheme can be achieved by writing:
\begin{align}\label{EQ:Spin_boson_EOM3}
   \S_{i+1} = \mathrm{e}^{\bm{\Omega}(R_{i+1},\dot{R}_{i+1})\delta t/2}\cdot\mathrm{e}^{\bm{\Omega}(R_i,\dot{R}_i)\delta t/2}\cdot\S_{i}
\end{align}
where 
\begin{equation}
     \bm{\Omega}(R,\dot{R})={\scriptstyle\left(\begin{array}{ccc}
       0 &  -\frac{2}{\hbar}V_z(R) & 2d(R)\dot{R}\\
        \frac{2}{\hbar}V_z(R) & 0 & 0\\
        -2d(R)\dot{R} & 0 & 0
    \end{array}\right)}.
\end{equation}\normalsize
In cases where a hop takes place, the whole step is split into smaller steps to determine the exact hopping place such that it is clear which active state to use for the force at each time. 
This ensures that the trajectories are exactly time-reversible.\cite{MASHEOM}
Note that the integrator can alternatively be defined in terms of wavefunction overlaps instead of nonadiabatic coupling vectors.


As discussed in \refsec{SEC:TPS}, it is crucial to define the boundaries of the reactant and product states before the application of MASH-TPS. 
For nonadiabatic systems, one has a choice as to whether to distinguish different adiabatic surfaces in the definition of the two states or solely utilize a position-based definition by bundling them together. In our case, we found that this distinction does not significantly affect the result, 
as the upper adiabatic state is unlikely to be populated in the reactant or product wells.
As such, 
the reactant is identified by all the points that satisfy $R<-5$ and $S_z<0$ (i.e., restricted to the ground state only) and the product by those with $R>5$ (irrespective of the value of $S_z$). 

\begin{figure}[t]
    \centering
    \includegraphics[width=1\linewidth]{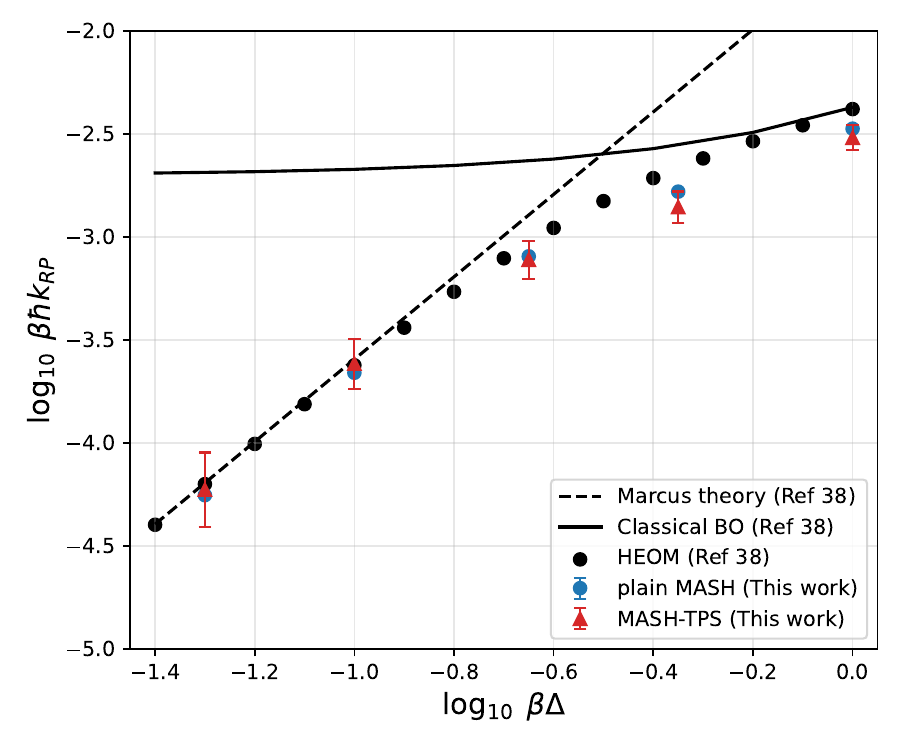}
    \caption{The reaction rate constant of spin--boson models with varying diabatic couplings, calculated through brute-force (blue triangles) and MASH-TPS (red circles) experiments. The error bars show $2\sigma$ confidence interval of the MASH-TPS rates, whereas the direct MASH simulations are fully converged. Both sets of experiments have quantitatively good overall agreement. The values for Classical BO, Marcus theory and HEOM (taken from Ref.~\onlinecite{MASHrates}) are included for comparison.} 
    \label{FIG:SB_rate_constant}
\end{figure}

After this definition, we proceed to calculate the rate constant in two ways. First, we perform a brute-force calculation by thermally sampling 5 ensembles of $8\times10^5$ phase-space points from the reactant region using an MC sampler. The spin vector $\S$ for each point is uniformly sampled from the lower half of the Bloch sphere.
In this way, we generate the classical Boltzmann ensemble of the ground state $\rho_0\propto\exp(-\beta E(\q,\p,\S))h(-S_z)$. 
Finally, the rate constants $k_{\stateR\stateP}$ for this portion of our experiments were determined as a time-average over the range in which $k(t)$ is constant [\reffig{FIG:SB_plain_ct_kt} in the supplementary material].
The final results 
are reported as blue circles in \reffig{FIG:SB_rate_constant} after averaging over the 5 ensembles.
This result will be used as the benchmark to test our implementation of the MASH-TPS procedure.
\begin{figure}[t]
    \centering
    \includegraphics[width=1\linewidth]{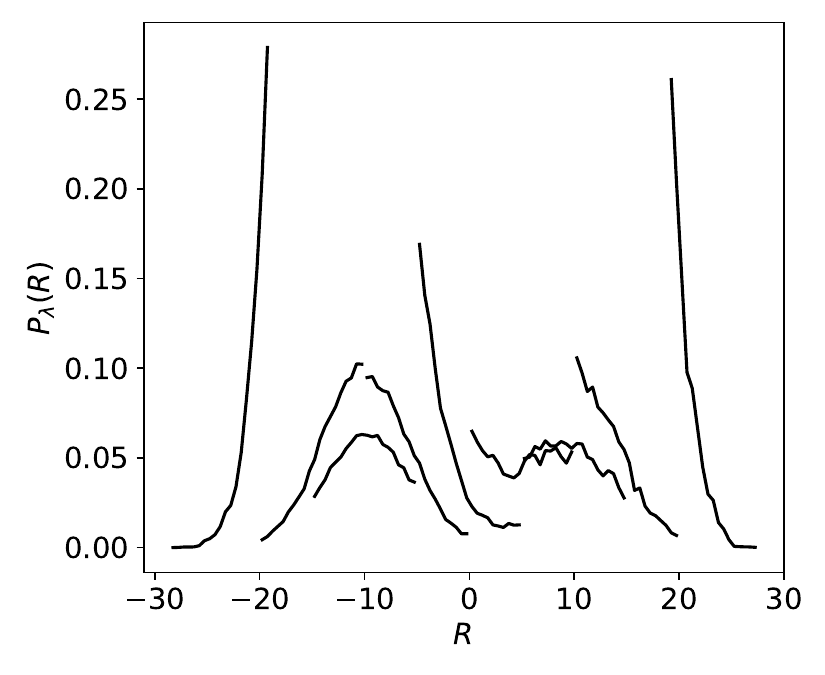}
    \caption{An example of the unweighted probability distributions for the 9 overlapping regions, $\lambda$, for the spin--boson model with $\beta\Delta=1$, obtained with the procedure outlined in \refsec{SEC:rate_const}. 
    }
    \label{FIG:SB_disconnected_Ctprime}
\end{figure}
\begin{figure}[t]
    \centering
    \includegraphics[width=1\linewidth]{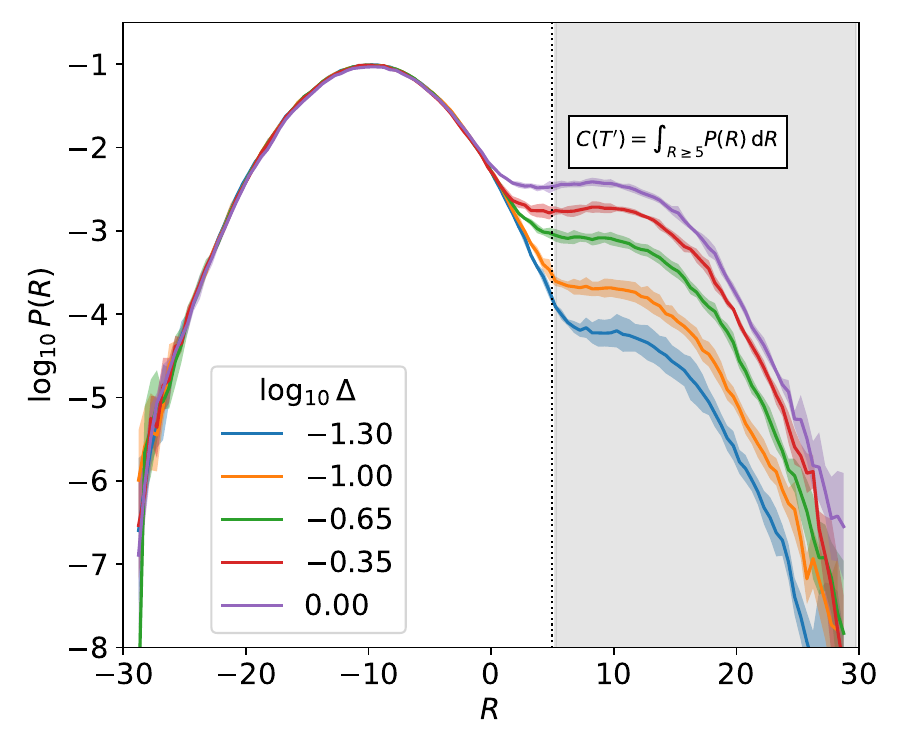}
    \caption{The graph of (connected) probability distribution averaged over six separate experiments where we first performed MASH-TPS calculations for the 9 overlapping regions (as the example shown in \reffig{FIG:SB_disconnected_Ctprime}) and then reconstructed with the WHAM, as outlined in \refsec{SEC:rate_const}. The colors correspond to different values of $\Delta$. The shaded area around each distribution represents a $2\sigma$ error estimate after the averaging. As indicated in the figure, the value of $C(T')$ is obtained by numerically integrating the mean distributions in the product state.}
    \label{FIG:SB_ct_prime_all}
\end{figure}
\begin{figure*}[t]
    \centering
    \includegraphics[width=1\linewidth]{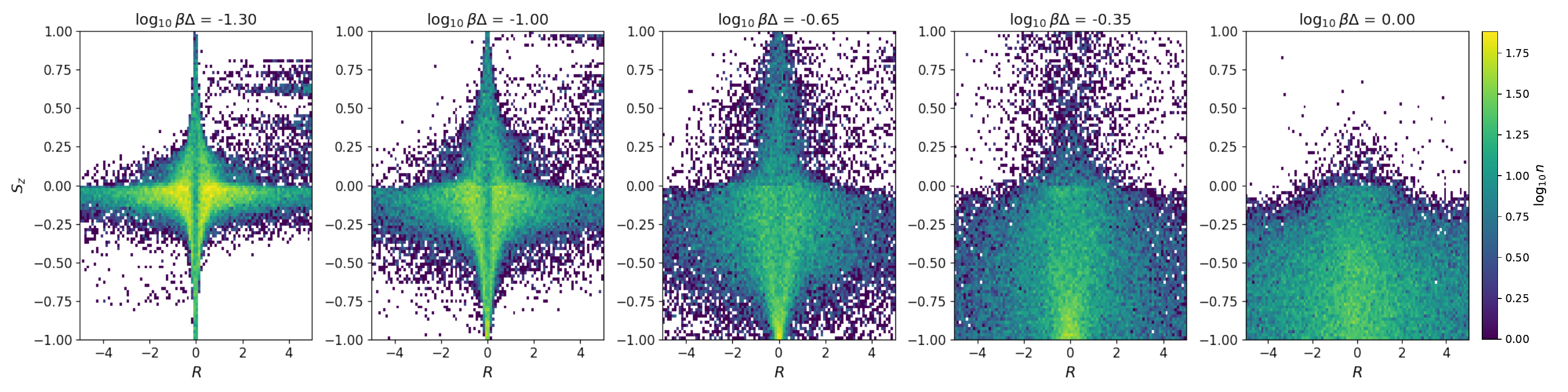}
    \caption{The distribution of accepted shooting moves in the transition-path ensemble of spin--boson models with a range of diabatic couplings.} 
    \label{FIG:SB_shooting_accepted}
\end{figure*}

Next, we evaluate the rate using MASH-TPS. From the plain MASH ensembles and for each value of $\Delta$, we randomly select 5  trajectories of length $T=50\beta\hbar$ 
which obey the reactivity condition $h_\stateR(0)=h_\stateP(T)=1$. From these, we initialize 5 independent Markov Chain MC processes where $2\times10^5$ trial path generation steps are performed with a 4--to--1 ratio between shifting (with shifts of $L=100$ steps in either direction) and alternating momentum and spin shooting moves (with 
$\delta \dot{R}$ drawn from a normal distribution with variance $0.01\beta^{-1}$ and $\chi$ drawn from a normal distribution with variance $0.1$\,radians).
These parameters were chosen to ensure a reasonable efficiency ($\approx 20\%$ across the range of diabatic couplings) in accepting new trial moves.
From the resulting TPS ensemble, one can generate the ratio in \refeq{EQ:final_Ct} for any value of $t$ and $T'$ in the range $[0,T]$.


Subsequently, for the free-energy calculation necessary to recover the factor of $C(T')$ from \refeq{EQ:final_Ct}, 
we first divide the space of the polarization coordinate $R$ into 9 separate overlapping regions, indexed by $\lambda$, covering the interval $R\in[-30,30]$. Then, similar to the previous step, we randomly selected $5$ initial trajectories of length $T'=20\beta\hbar$, corresponding to each region. 
These trajectories must satisfy $h_\stateR(0)=h_{\lambda}({T'})=1$, with $h_\lambda$ being the characteristic function of the region. Finally, we carry out $2\times10^{5}$ MC steps starting from each initial trajectory and histogram the resulting values of $R(T')$. An example of disconnected distributions, $P_\lambda(R)$, obtained from this procedure can be seen in \reffig{FIG:SB_disconnected_Ctprime}. The final probability distribution, $P(R)$, reported in \reffig{FIG:SB_ct_prime_all}, is constructed via first performing the WHAM procedure for each ensemble, and then averaging over the 5 independent ensembles. The value of $C(T')$ is evaluated by numerically integrating this distribution over the product region $R\geq5$.
Finally, $C(t)$ can be constructed using \refeq{EQ:final_Ct}, for which plots are included in the supplementary material\@.
Similar to the brute-force calculation above, we determine the rate constant $k_{\stateR\stateP}$ by taking the time average of $k(t)=\mathrm{d}C(t)/\mathrm{d}t$ in the range where it is constant (\reffig{FIG:SB_MTPS_ct_kt} in supplementary material). 
The whole calculation is carried out on 5  different models with varying values of $\beta\Delta$. 
All the simulation parameters pertaining to both sets of experiments are summarized in \reftable{TABLE:SB_Sim_details}\@. 
\begin{table*}[!t]
    \caption{Simulation parameters of the MASH and MASH-TPS calculations for the spin--boson model for each value of the diabatic coupling $\Delta$ reported in \reffig{FIG:SB_rate_constant}. The cumulative simulation time of MASH-TPS accounts for the fact that, for every shooting move, 4 shifting moves with $L=100$ shifting steps were performed. The assumption here is that for every shooting move, the forward and backward segments of the trajectory were completed before the reactivity conditions were checked. Such an assumption gives the maximum simulation time for the MASH-TPS calculations.}
    \centering
    \begin{tabular}{l|c|c|c|l}\label{TABLE:SB_Sim_details}
        Calculations & Num. of ensembles & Num. of traj. per ensemble & Duration of traj. 
        & Cumulative num. of steps\\\hline
        {\scriptsize } & &  &  &\\
        MASH & 5 & $N_\text{MASH} = 8.0\times10^{5}$ & $200\beta\hbar=20\,000\delta t$ & $8.\times10^{10}$\\
        MASH-TPS & $5$ & $N_\text{TPS} = 2\times10^{5}$ & $T=50\beta\hbar=5000\delta t$ & $1.08\times10^9$\\
        $C(T')$ [9 regions] & $5\times9=45$ & $N'_\text{TPS} = 2\times10^{5}$ & $T'=20\beta\hbar=2000\delta t$ & $3.74\times10^{9}$
    \end{tabular}
\end{table*}
Most importantly, \reffig{FIG:SB_rate_constant} shows that the MASH-TPS results quantitatively agree with the plain MASH simulations (within error bars), which confirms the correctness of the MASH-TPS procedure.
In this particular case, there is no dramatic efficiency gain from using MASH-TPS rather than the brute-force approach.
Although we computed the MASH-TPS rates using fewer and shorter trajectories, in hindsight, one could clearly reduce the length and number of the trajectories in the the brute-force simulation without significantly impacting the error bars.
However, this behavior is of course strongly system-specific.
We chose this particular system to have a low enough barrier such that the brute-force simulation could be converged and provide a benchmark.
It is expected that MASH-TPS will give a significant efficiency gain for systems with much larger barriers, e.g., by increasing the reorganization energy, $\Lambda$, for which the error of the brute-force simulation will increase exponentially.

After this discussion of the efficiency, it is worth reminding the reader of the most important aspect of computer simulation.
Only MASH gives a reliable rate constant (whether evaluated directly or using the TPS approach), whereas FSSH underestimates the rates in the nonadiabatic limit, as was shown in Ref.~\onlinecite{MASHrates}.

Finally, we briefly discuss how TPS can be used to characterize the reaction mechanism. 
The conditional probability $p(\text{TP}|\x)$ quantifies the probability for a pathway passing through $\x$ to be reactive. 
To parametrize $p(\text{TP}|\x)$ using a TPS ensemble, Peters and Trout \cite{Peters2006aimless} provide a method in which the parameters of a postulated model are
adapted to maximize the likelihood of observing a particular sequence of accepted and
rejected shooting moves. Extensions to include nonlinear dependence on CVs and/or using Artificial Neural Networks also exist in the literature.\cite{Peters2006aimless,Lechner2010,Jung2023} Our aim in this paper, however, is not to build an explicit model for $p(\text{TP}|\x)$. Rather, motivated by the formalism adopted in Ref.~\onlinecite{Peters2006aimless},  \reffig{FIG:SB_shooting_accepted} shows histograms of accepted shooting moves in the space spanned by $R$ and $S_z$ for each diabatic coupling.
These plots are indicators of $p(\text{TP}|\x)$ in the transition region.

It is evident from the figure that, 
for strong diabatic coupling, nearly all reactive transitions occur on the ground state and that the closer the shooting point is to $R=0$, the higher the probability of generating a reactive trajectory.
This follows the expectation that for an adiabatic reaction, the optimal transition state is located at the barrier top.
However, for weak diabatic coupling, most reactive trajectories have spin vectors just below the Bloch-sphere equator, where they have a higher probability of remaining on the ground state after passing through the region of strong nonadiabatic coupling (which almost inverts the spin vector\cite{MASHrates}).
In this case, there is also a reasonable probability of generating reactive trajectories from the excited state, but this is limited to the low-energy region around the origin. 
Additionally, there is a spike at $R=0$ where 
the trajectories are midway through the nonadiabatic coupling region, meaning that the spin vector can rotate significantly in order to reach (or remain on) the ground state in both the forwards and backwards directions.
This dramatic shift in the behavior of the histograms captures the change in mechanism from adiabatic to nonadiabatic reactions and can be used to characterize the importance of nonadiabatic effects.


\section{Conclusion and Discussion}
In this paper, we have developed an efficient molecular dynamics framework based on a combination of MASH and TPS methods to study nonadiabatic reactions. MASH-TPS is particularly useful for complex reactions that exhibit a separation of timescales, but for which a reaction coordinate is not at our disposal \emph{a priori}. The key ingredient that allowed for its development was the Markovianity and time reversibility provided by MASH trajectories. In this way, the MASH formalism is not just more accurate than FSSH but also provides a more efficient approach to simulate nonadiabatic dynamics. Moreover, in reactions hampered by entropic effects, the lack of requirement for a reaction coordinate makes TPS more robust in comparison to other path sampling methods like forward-flux sampling. The efficiency of TPS is particularly prominent when one only needs to identify the reaction mechanism from the ensemble and can skip the calculation of $C(T')$. 

In our application, we restricted ourselves to a system that could be sufficiently described by two adiabatic states. 
However, the framework introduced here is also directly applicable to multi-state cases using multi-state extensions of MASH.\cite{Runeson2023MASH,Runeson2024MASH,Runeson2024semiconductors,runeson2025nuclear,unSMASH,Granucci2025MASH,MASHreview} In particular, we recommend
the unSMASH approach of Ref.~\onlinecite{unSMASH}
for photochemistry applications, whereas for exciton models, the methods of Refs.~\onlinecite{Runeson2023MASH} and \onlinecite{Runeson2024MASH} are preferred. 


With regard to the TPS component of our framework, we note that more sophisticated shooting strategies have been developed for the generation of trial trajectories.
For instance, specialized methods exist for guiding the noise in trajectories with stochastic forces.\cite{Gingrich2015preserving}
Another method called `shooting-from-the-top'\cite{Jung2017} restricts shooting to configurations in the vicinity of the (predefined) transition state, and trajectory propagation is terminated once the path reaches the boundaries of the metastable basins. This strategy significantly enhances sampling efficiency by focusing computational effort on the transition region and avoiding unnecessary sampling within the metastable states. The statistical quantities, such as rate constants, can be retrieved in a somewhat similar fashion to the original TPS method outlined above. In a closely related approach, rather than prescribing the shooting region \emph{a priori}, Jung et al.~\cite{Jung2023} employ a scheme where a neural network adaptively identifies the transition state based on information obtained from previously accepted and rejected shooting points. This data-driven strategy enables a dynamic localization of the transition region during the sampling procedure.


A particular advantage of the TPS formalism is that it gives mechanistic insight into the reaction process.
In particular, it gives easy access to the committor, which quantifies the likelihood of a trajectory initiated from a configuration reaching the product state before returning to the reactant, and is therefore commonly regarded as the ideal reaction coordinate.\cite{Dellago2006TPS,Bolhuis2015} 
In fact, its maximum in the transition region coincides with the true transition state. 
We have shown how this approach identifies the bottlenecks of the reaction mechanism and gives insight into the role played by nonadiabatic transitions.

Finally, we also briefly highlight three cases recently studied in our group 
where MASH-TPS could be a valuable tool.
In photodissociation, one is often interested in the yield of the various products.
This can in principle be obtained from brute-force simulations,\cite{cyclobutanone}
although one then obtains poor statistics on the minor product channels.
In principle, MASH-TPS would allow the study of the rates, mechanisms and relative yields of different photodissociation products.

In another study, MASH was employed to predict the electronic coherences after the systems passes near a conical intersection.\cite{MASHcoh} It was shown that MASH accurately reproduces these coherence observables, demonstrating its potential as a reliable simulation approach for novel X-ray spectroscopies.
However, more trajectories were required to converge the coherences than for standard simulations of time-dependent populations.
We propose that the MASH-TPS framework can be leveraged to enhance the sampling of coherences by focusing on trajectories which spend time near the equator where the contribution is largest.

Finally, we have recently introduced a formalism to study open quantum systems, by combining the deterministic dynamics of MASH with a stochastic unravelling of secular Redfield theory.\cite{KasraOpenQuSys}
With this method, we showed that one can capture spontaneous emission from a molecule undergoing nonadiabatic processes.  However, as spontaneous emission is a rare event, many trajectories were required to obtain good statistics on the photon emission probabilities.
As the dynamics is sill Markovian, we propose that such processes could be more efficiently studied using MASH-TPS.

\section*{Supplementary Material}

\begin{figure*}[h]
    \centering
    \includegraphics[width=0.9\textwidth]{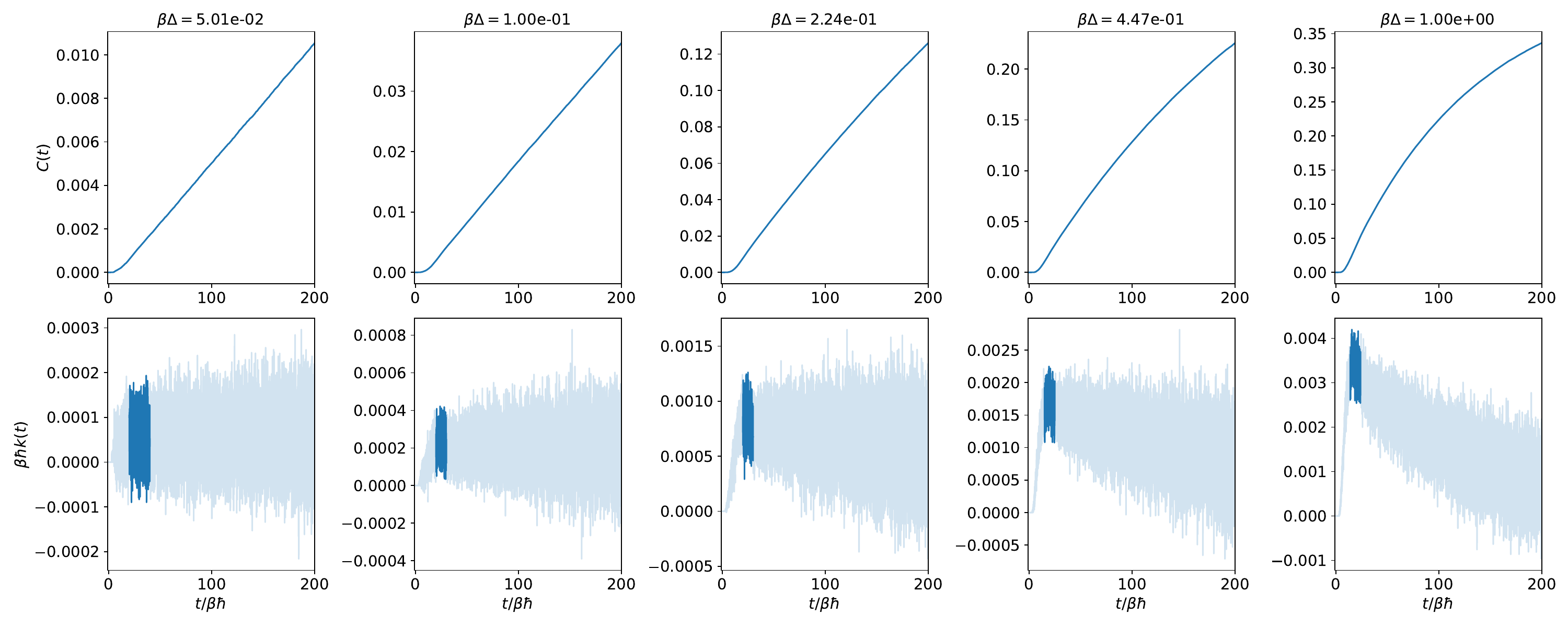}
    \caption{The plot of $C(t)$ (top) and $k(t)$ (bottom) for plain MASH ensembles in the spin--boson models. The range in $k(t)$ graphs from where we calculate the brute-force rate constants (Fig. 4 of the main text) is highlighted for each value of $\beta\Delta$.}
    \label{FIG:SB_plain_ct_kt}
\end{figure*}

In \reffig{FIG:SB_plain_ct_kt}, we have plotted $C(t)$ and $k(t)$ for the spin--boson model calculated from an ensemble of plain MASH trajectories for {5} values of diabatic coupling. The period in which $k(t)$ becomes constant for each $\Delta$ has been highlighted. It is worth mentioning that as the value of diabatic coupling $\Delta$ increases, the nonadiabatic effects dampen, and with it the rarity of transition. This results in a shift of the constant period of $k^{\text{plain}}(t)$ toward zero and also shrinking of its size. 

\begin{figure*}[h]
    \centering
    \includegraphics[width=0.7\textwidth]{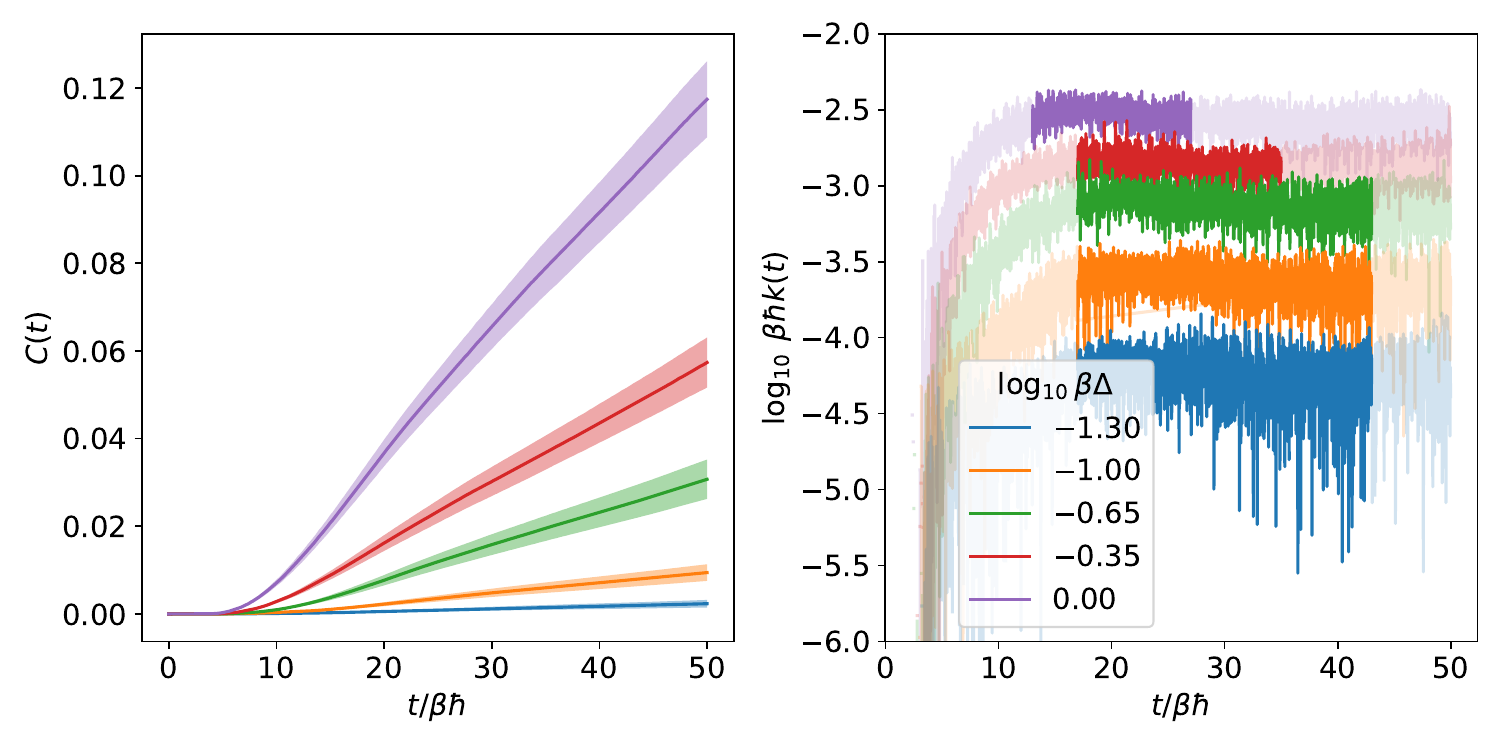}
    \caption{The plot of normalized $C(t)$ (left) and $k(t)$ (right) calculated from $5$ individual MASH-TPS ensembles. Similar to \reffig{FIG:SB_plain_ct_kt}, we have highlighted the period for calculating the rate constant in the $k(t)$ graph. The shaded areas depict the $2\sigma$ error of averaging over 5 ensembles.}
    \label{FIG:SB_MTPS_ct_kt}
\end{figure*}

Figure \ref{FIG:SB_MTPS_ct_kt} shows the same quantities calculated with MASH-TPS and averaged over 5 ensembles. 

\section*{Author Contributions}

\bibliography{bibliography,references}

\clearpage


\end{document}